\documentclass[
    ,final            
  ]
  {aipproc}

\layoutstyle{6x9}

\begin{document}

\title{HADRON05 Summary:\\ Heavy Quark Hadrons and Theory}


\author{T.Barnes}{
  address={Physics Division, ORNL \\
and\\
Department of Physics and Astronomy,\\
University of Tennessee}
}

\begin{abstract}
This HADRON05 summary covers the topics of (1) mesons containing heavy quarks, 
and (2) theory. The new material discussed here is taken mainly 
from plenary presentations. We specifically emphasize 
new or recent results in spectroscopy that are likely to appear 
in future editions of the PDG. An exception is 
made for the pentaquark, which has now been withdrawn. We undoubtedly
have something important to (re)learn about multiquarks 
from the pentaquark saga, and this merits a phrase in Portuguese.
The three general areas we consider are: I. $Q\bar Q$ spectroscopy,  
II. $Q\bar q$ spectroscopy, and III. lessons from the pentaquark.
Finally, in Section IIIb. we conclude with ``Our moment of Zen''.
\end{abstract}

\maketitle


\section{$Q\bar Q$ spectroscopy}

\subsection{Introduction}

The most exciting new results presented at HADRON05 (excepting the withdrawal
of the pentaquark) were the new experimental developments in 
charmonium spectroscopy. After a hiatus of approximately 20 years, 
there have been many exciting new results in the charmonium 
sector, due largely to the rather surprising effectiveness of B decays 
in producing $c \bar c$ resonances \cite{Eichten:2002qv}.
This was discussed here 
by Trabelsi \cite{Trabelsi}, 
who noted that the CKM-favored B meson weak decay process  
leads to a relatively large ({\it ca.} 1\%) inclusive $c\bar c$ 
branching fraction, due to the favorable $V_{bc}$ and $V_{cs}$ 
CKM matrix elements in the 
quark-level process $b \to c \bar c s$. The $c\bar c$ mesons identified in 
B decays to date include not only the S-wave J$/\psi$, $\eta_c$, $\psi'$ 
and $\eta_c'$, but also the P-wave states $\chi_0$ and $\chi_1$ and the 
dominantly D-wave $\psi(3770)$.
This is a surprising result, since the W propagator in the diagram might 
{\it a priori} have been expected to couple preferentially to a final $c\bar c$
system at small separations of $\approx 1/M_w \approx 0.01 fm$. 
Production of P-wave and dominantly D-wave $c\bar c$ states with branching 
fractions comparable to those of S-wave states has nonetheless been reported
experimentally. This gives us an effective method for producing higher-mass 
radially- and orbitally-excited $c\bar c$ states, and perhaps other states in
this flavor sector, such as $c\bar c$ hybrids or charm molecules. 

Four new states which might be charmonia were discussed at 
this meeting in plenary presentations; these are the Z(3931), Y(3943), 
X(3943) and Y(4260). (The earlier discovery of the X(3872) has apparently 
led to a preference for the far end of the Latin alphabet in naming these 
new states.) 
We will discuss the new states individually, and review the possibilities 
discussed for assignments in the conventional $c\bar c$ spectrum.

\subsection{Z(3931)}

The new states Z(3931), Y(3943) and X(3943) reported by Belle 
all have possible assignments as conventional 2P $c\bar c$ states; to misquote
Hamlet, ``2P or not 2P, that is the question''.
The Z(3931), discussed here by Trabelsi \cite{Trabelsi} was reported
in $\gamma \gamma \to Z \to DD$ (both charge
states \cite{Abe:2005bp}; see figure~1). It has a reported mass 
and width of

\begin{equation} 
{\rm M} = 3931 \pm 4 \pm 2 \ {\rm MeV},
\label{Z3931_mass}
\end{equation}
\begin{equation} 
\Gamma = 20 \pm 8 \pm 3 \ {\rm MeV}.
\label{Z3931_width}
\end{equation}
Belle suggests a 2P $\chi_2'$ assignment for the Z(3931). 
This remarkable state has a surprisingly small total width of about 20
MeV, which is about 1/2 of what is expected for a 2P  $\chi_2'$ $c\bar c$ 
meson at this mass \cite{Barnes:2005pb}.
The reported two-photon width is 
\begin{equation}
\Gamma_{\gamma\gamma} \cdot B_{DD} = 0.23 \pm 0.06 \pm 0.04 \ {\rm keV},
\label{Z3931_twophoton_width}
\end{equation}
which is about 1/2 the theoretical 
expectation for a $\chi_2'$ $c\bar c$ state \cite{Bar92} (neither the total
width nor the two-photon width is a disaster, given the typical accuracy of 
these calculations). 

Another possibility is that the Z(3931) might be a 
2P $\chi_0'$ $c\bar c$ scalar; decay calculations using the ${}^3$P$_0$ model
show that this state should be surprisingly narrow, due to a node
in the DD decay mode which is accidentally near 3.95 GeV. 
One can differentiate between these options by searching for DD and DD$^*$ 
modes; the scalar $\chi_0'$ will only produce DD, whereas the tensor 
should decay to both DD and DD$^*$, with an expected branching fraction 
ratio of DD$^*$/DD $ \approx 1/3$. 

\begin{figure}[ht]
\includegraphics[height=.3\textheight]{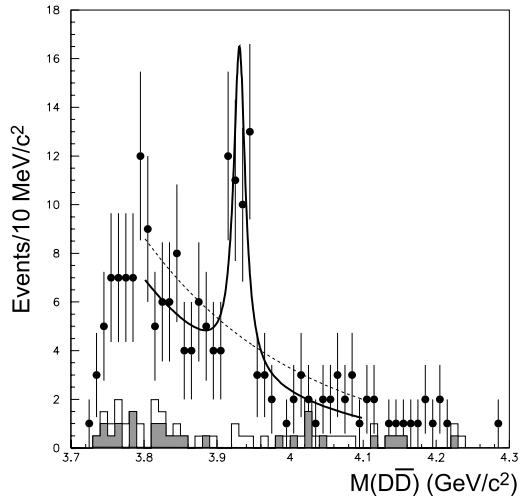}
\caption{The Z(3931) reported in $\gamma\gamma \to $Z$ \to $DD
by Belle \cite{Abe:2005bp}.}
\end{figure}

\subsection{Y(3943)}

The Y(3943) was reported in B decays to KY, Y$\to  \omega J/\psi$ 
\cite{Abe:2004zs}, with a mass and width of
\begin{equation}
{\rm M} = 3943 \pm 11 \pm 13 \ {\rm MeV},
\label{Y3943_mass}
\end{equation}
\begin{equation} 
\Gamma = 87 \pm 22 \pm 26 \ {\rm MeV}.
\label{Y3943_width}
\end{equation}

The combined Y(3943) formation and decay branching fraction found by Belle is
\begin{equation}
B_{{\rm B}\to {\rm KY(3943)}} \cdot
B_{{\rm Y(3943)}\to \omega J/\psi }
= 7.1 \pm 1.3 \pm 3.1 \cdot 10^{-5}.
\label{Y3943_BFs}
\end{equation}
Since a typical total branching fraction for production of conventional
charmonium states in B decays is $10^{-3}$ to $10^{-4}$, this suggests that
the Y(3943) has a large branching fraction to the closed charm channel 
$\omega J/\psi$, if this enhancement is indeed due to a real resonance. 

The remarkable closed-charm $\omega J/\psi$ mode led Belle to suggest that 
the Y(3943) might be a charmonium hybrid \cite{Abe:2004zs}. 
They note however that the mass of the Y(3943) is 500 MeV below LGT estimates;
this makes the hybrid assignment appear implausible. In view of
the mass of the Y(3943), another possibility which should be considered is
that this is another conventional 2P $c\bar c$ state; this would lead
to a total width comparable to what is reported for the Y(3943), with 
DD$^*$ as the dominant decay model. A search for this DD* mode is underway
\cite{Abe:2004zs}, and if found, this may lead to identification of this 
rather wide enhancement with the 2P $\chi_1'$  $c\bar c$ state. 

\begin{figure}[ht]
\includegraphics[height=.3\textheight]{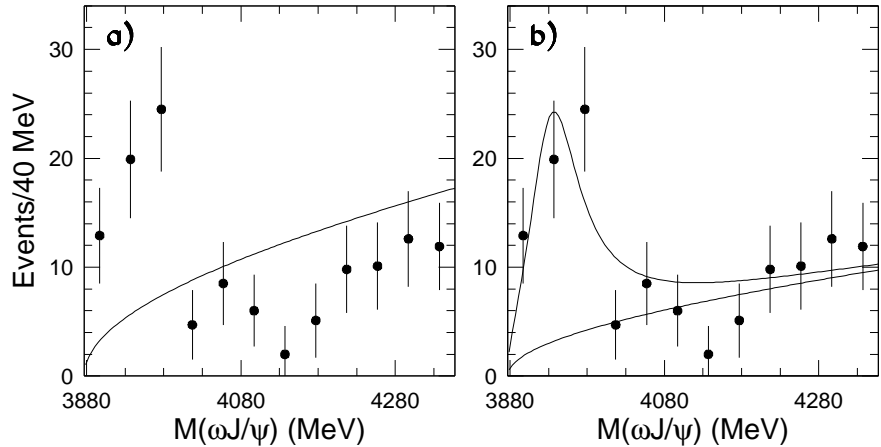}
\caption{The Y(3943) reported by Belle in the closed-charm mode
B$\to$ KY, Y$\to  \omega J/\psi$ (fitted in right panel)
\cite{Abe:2004zs}.}
\end{figure}

\subsection{X(3943)}

The X(3943) was reported by Belle in the double charmonium reaction
$e^+e^- \to J/\psi X$ ($X$ is mainly DD$^*$) \cite{Abe:2005hd}.
The fitted parameters are
\begin{equation}
{\rm M} = 3943 \pm 6 \pm 6 \ {\rm MeV},
\label{X3943_mass}
\end{equation}
\begin{equation} 
\Gamma = 15.1 \pm 10.1 \ {\rm MeV}.
\label{X3943_width}
\end{equation}

The X(3943) has an estimated total width of only about 15 MeV, however 
the missing mass bump in the 3940 region may have a more
complicated structure (see figure~3); 
this small width estimate evidently carries considerable uncertainty.

\begin{figure}[h]
\includegraphics[height=.3\textheight]{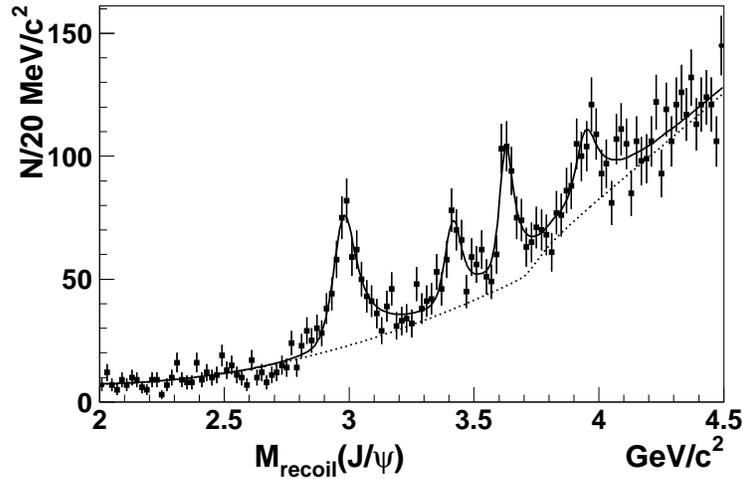}
\caption{The X(3943) reported by Belle in the double charmonium 
production reaction $e^+e^- \to J/\psi X$. The $\eta_c$, $\chi_0$
and $\eta_c'$ are also evident \cite{Abe:2005hd}.}
\end{figure}

Belle report that DD$^*$ is the dominant X(3943) decay mode,
with a branching fraction of
\begin{equation}
B_{{\rm DD}^*} = 96 {+45 \atop -32} \pm 22 \% .
\label{X3943_BF1}
\end{equation}
The other available open-charm mode is DD, which is only known as
an upper limit,
\begin{equation}
B_{{\rm DD}} <  41 \% , \ \ 90\% \ c.l.
\label{X3943_BF2}
\end{equation}
Similarly the closed-charm mode $ \omega  J/\psi$ (of interest since this
mode is apparently preferred by the Y(3943)) is also known as an upper limit,
\begin{equation}
B_{ \omega  J/\psi} <  26 \% , \ \ 90\% \ c.l.
\label{X3943_BF3}
\end{equation}
Eichten has suggested a 3$^1$S$_0$ $\eta_c$ $c\bar c$ radial excitation 
assignment for this state; given the 3$^3$S$_1$ $\psi(4040)$, the mass is
roughly correct, and the observation of the $\eta_c$ and $\eta_c'$ with 
comparable strength in double charmonium production supports this 
identification. The other possibility with a dominant DD$^*$ mode in this 
mass region is the 2$^3$P$_1$ $c\bar c$, but the absence of a strong 
signal for the 1$^3$P$_1$ $\chi_1(3510)$ in double charmonium production argues
against this 2P assignment. The 3$^1$S$_0$ $\eta_c$ assignment could be tested
by searching for evidence of this state in $\gamma\gamma \to $DD$^*$.

\subsection{Y(4260): A charmonium hybrid?}

The most remarkable new state discussed at HADRON05 was the Y(4260), 
which was reported by BaBar in initial state radiation (ISR) in the reaction
$e^+e^- \to \gamma_{ISR} J/\psi \pi^+\pi^-$ \cite{Aubert:2005rm}.
There may also be evidence for an enhancement in $J/\psi \pi^+\pi^-$
states near 4.26~GeV  
in the decay B$\to$K$J/\psi \pi^+\pi^-$ (in both neutral and negative
B/K charge states) \cite{Aubert:2005zh}. The mass and width reported for 
the Y(4260) \cite{Aubert:2005rm} are
\begin{equation}
{\rm M} = 4259 \pm 8 {+2 \atop -6} \ {\rm MeV},
\label{Y4260_mass}
\end{equation}
\begin{equation} 
\Gamma = 88 \pm 23 {+6 \atop -4} \ {\rm MeV}.
\label{Y4260_width}
\end{equation}

\begin{figure}[h]
\includegraphics[height=.3\textheight]{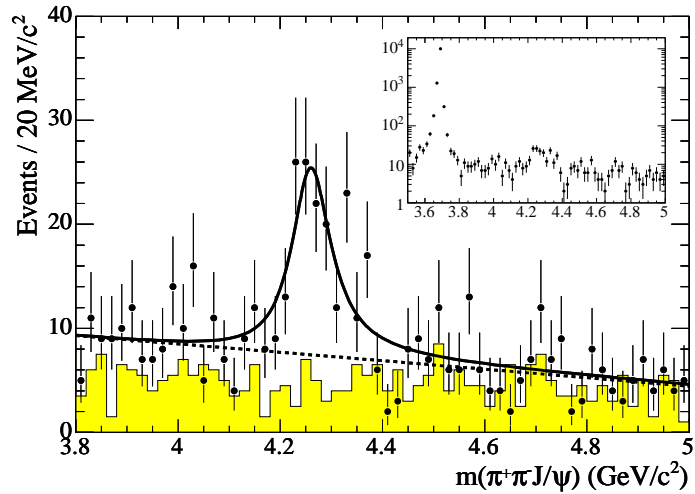}
\caption{The Y(4260) reported by BaBar in the ISR process
$e^+e^- \to \gamma_{ISR} J/\psi \pi^+\pi^-$ \cite{Aubert:2005rm}.}
\end{figure}

The ISR production mechanism tells us that this state must be $1^{--}$,
but it cannot be a conventional $c\bar c$ state because the $1^{--}$
states in this mass region are well established from earlier $e^+e^-$
annihilation experiments. (The Y(4260) is bracketed by the 
2$^3$D$_1$ $\psi(4160)$ and the 4$^3$S$_1$ $\psi(4415)$, which have
masses that are in excellent agreement with the expectations of
$c\bar c$ potential models.) 

The Y(4260) (if a real resonance) evidently represents ``overpopulation'' 
of the expected quark potential model spectrum of $1^{--}$ $c\bar c$ states.
This observation, combined with the fact that there is no enhancement
visible in R near this mass \cite{Seth} has led to suggestions that this 
state may be a charmonium hybrid \cite{Close:2005iz}.
The lightest hybrid multiplet in both the flux tube and bag models
contains a $1^{--}$ state. In the flux tube model this lightest $c\bar c$-hybrid
multiplet is predicted to lie near 4.2~GeV \cite{Barnes:1995hc}, and 
in LGT studies the lightest $1^{-+}$ exotic hybrid (expected to be near the 
$1^{--}$ hybrid in mass) is found at a rather similar mass of
about 4.4~GeV \cite{Liao:2002rj}.
In these models hybrids have a vanishing $c\bar c$ 
wavefunction at contact, so it has long been speculated that they would 
have small $e^+e^-$ widths, and thus make rather weak contributions to R. 
A LGT study by the UKQCD group \cite{McNeile:2002az}
of the strong decay couplings of exotic $b\bar b$ hybrids found strikingly 
large couplings to closed flavor modes (specifically to $\chi_b S$, 
where $S$ is a light scalar isoscalar meson that would decay to $\pi\pi$).
This is sufficiently similar to the BaBar observation of 
the Y(4260) in the single closed-charm mode $J/\psi \pi^+\pi^-$ to be
cited as additional possible evidence for a hybrid interpretation. 

The unusual $J/\psi \pi^+\pi^-$ mode and the UKQCD study
suggest searches of any other accessible closed-charm
modes with $1^{--}$ quantum numbers, such as 
$J/\psi \eta$, $J/\psi \eta'$, $\chi_J \omega$ and
so forth. Ideally the light system should have quantum numbers 
thought to couple strongly to pure glue, such as $0^{++}$ and $0^{-+}$.

In specific decay models (flux tube and constituent glue), theorists 
have anticipated that the dominant open-charm decay modes of charmonium hybrids
would be a meson pair with one internal S-wave (D, D$^*$, D$_s$, D$_s^*$) 
and one internal P-wave (such as D$_J$ and D$_{sJ}$). 
In the case of the Y(4260) this suggests dominance of the decay mode
DD$_1$(2430). This broad D$_1$ has a width of {\it ca.}~400~MeV, and decays
to D$^*\pi$, so this suggests a search for evidence of the Y(4260) in 
DD$^*\pi$.
Since this is a prediction of a decay model in an untested regime (hybrids),
one should be cautious and search the more familiar two-body modes
DD, 
DD$^*$, 
D$^*$D$^*$,
D$_s$D$_s$,
D$_s$D$_s^*$
and
D$_s^*$D$_s^*$
for evidence of the Y(4260) as well.
If there is evidence of a large DD$_1$(2430) signal, the Y(4260) would then be 
quite convincing as a hybrid having properties predicted by the flux tube 
model. If it appears in some of these open charm modes such as 
DD$^*$ and D$_s$D$_s^*$
at rates comparable to or larger than $J/\psi\pi^+\pi^-$, 
one might claim a hybrid 
but speculate that the flux tube decay model was inaccurate in predicting
hybrid decay modes. Finally, if the Y(4260) does not appear in any other mode,
one might be skeptical about whether the $J/\psi\pi^+\pi^-$ signal is due to a
resonance at all; there are nonresonant possibilities, such as 
production of DD$_1$ in
$e^+e^- \to $ DD$_1$ followed by an inelastic FSI that produces a broad 
$J/\psi\pi^+\pi^-$ enhancement due to the (very broadened) onset of 
DD$_1$(2430) threshold events (which would appear near 4.3~GeV).

In any case the evidence for the Y(4260) from BaBar is not strong, with 
an estimated $125\pm 23$ events \cite{Aubert:2005rm}. Since it is clear
from the reported  mass that the Y(4260) is not a conventional $c\bar c$ 
state, it will be very important to test the existence of this state 
through the accumulation of better statistics.

\subsection{Other developments in the $Q\bar Q$ sector: X(3872)}

Progress in other sectors of $Q\bar Q$ has been less dramatic, 
although there have been some interesting developments. The unusual and very
narrow X(3872) was reviewed here by Trabelsi \cite{Trabelsi}
and by Swanson \cite{Swanson}. There is now clear evidence that this state
is $1^{++}$, which was a prediction of the DD$^*$ molecule picture promoted
in particular by Tornqvist \cite{Tornqvist:2003na} and Swanson 
\cite{Swanson:2003tb}. New results from CDF~II discussed here by
Maciel \cite{Maciel} show that the $\pi^+\pi^-$ mass distribution in the 
decay mode X(3872)$\to J/\psi \pi^+\pi^-$ is well described by the 
assumption that it is due to $J/\psi \rho^0$. One especially striking prediction 
of the (neutral D) D$^0$D$^{*0}$ molecule model is that one should observe 
comparable strength $J/\psi \omega$ and $J/\psi \rho^0$ decay modes 
\cite{Swanson:2003tb}, 
due to the maximal isospin breaking present in the initial state.
This prediction appears to have been confirmed by the evidence from Belle
for the $\omega$ in the $3\pi$ mode
X(3872)$\to J/\psi \pi^+\pi^-\pi^0$ \cite{Abe:2005ix}. The $3\pi$ invariant 
mass peaks at the highest mass, as expected for a virtual $\omega$, and the 
ratio of $2\pi$ to $3\pi$ branching fractions is close to unity, 
\begin{equation}
\frac{\Gamma({\rm X(3872)} \to J/\psi \pi^+ \pi^- \pi^0)} 
{\Gamma({\rm X(3872)} \to J/\psi \pi^+ \pi^-)} = 1.0 \pm 0.4 \pm 0.3,
\end{equation}
as expected in the DD$^*$ model.
There is also evidence from Belle 
for the radiative transition X(3872)$\to \gamma\; J/\psi$ \cite{Abe:2005ix},
with the width ratio 
\begin{equation}
\frac{\Gamma({\rm X(3872)} \to  \gamma\; J/\psi)} 
{\Gamma({\rm X(3872)} \to J/\psi \pi^+ \pi^-)} = 0.14 \pm 0.05,
\label{X3872_rad}
\end{equation}
which should be useful in testing the details of different models of the
X(3872). 

\section{Heavy-light meson spectroscopy}

Most of the studies of heavy-light mesons reported at HADRON05 
considered decays of known $Q\bar q$ states, and these have 
provided very extensive new results on light meson
spectroscopy (see for example the talk by Asner \cite{Asner}). 
There has been rather less recent activity on the spectroscopy of heavy-light mesons
themselves. Several new results on $Q\bar q$ spectroscopy were presented 
at HADRON05, and I will specialize to a few of these in the 
B and D sectors. There are also some new results on heavy-quark baryons, such
as the identification by SELEX of the first doubly-charmed baryon, a
$\Xi^+_{cc}$ at $3518.7\pm 1.7$~MeV; baryon spectroscopy was discussed 
here by Cumulat \cite{Cumulat} and in the summary by Klempt \cite{Klempt}.  

Maciel \cite{Maciel} discussed results from CDF on the B$_c^{\pm}$ 
pseudoscalar meson, which is reported in decays to $J/\psi \pi^{\pm}$
\cite{Acosta:2005us}.
The fitted mass determined from the {\it ca.}~19~events was
\begin{equation}
{\rm M}_{{\rm B}_c} = 6287.0 \pm 4.8 \pm 1.1~{\rm MeV},
\label{Bc_mass}
\end{equation}
which is roughly consistent with or somewhat below 
theoretical expectations.  
\begin{figure}[h]
\includegraphics[height=.25\textheight]{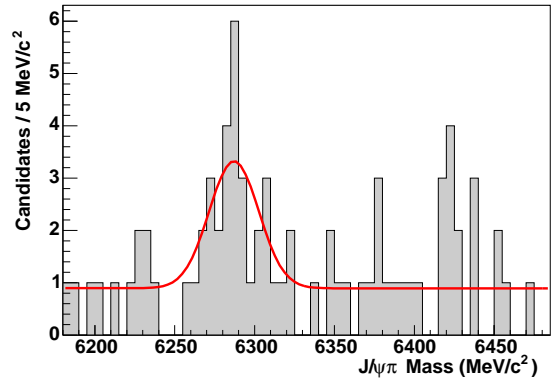}
\caption{The $B_c$ reported by CDF in $B_c^{\pm} \to J/\psi \pi^{\pm}$
\cite{Acosta:2005us}.}
\end{figure}

\eject
Results from CDF and D0 on orbitally excited B mesons were also presented;
for the first time we have evidence for the separate states in the P-wave
B meson multiplet, rather than seeing an unresolved superposition of states.
Both experiments identify B mesons in $J/\psi$K events, which are 
combined with another $\pi$ and studied for evidence of excited B mesons.
In both experiments the results are preliminary, but do show evidence of 
separate B$_2^*$ and B$_1$ contributions. The D0 mass results are
\begin{equation}
{\rm M}_{{\rm B}_1} = 5724 \pm 4 \pm 7~{\rm MeV},
\label{B1_mass}
\end{equation}
\begin{equation}
{\rm M}_{{\rm B}_2^*} - {\rm M}_{{\rm B}_1} = 23.6 \pm 7.7 \pm 3.9~{\rm MeV}.
\label{B2_mass}
\end{equation}
Their B$\pi$ mass plot is shown in figure~6.
The total widths, constrained in the fit to be equal (and predicted to be 
comparable theoretically) are
\begin{equation}
\Gamma_{tot.} = 23 \pm 12~{\rm MeV},
\label{B_width}
\end{equation}
which is consistent with theoretical expectations for the two narrow P-wave
B mesons. The experimental resolution is estimated to be about 10~MeV.

\begin{figure}[ht]
\includegraphics[height=.25\textheight]{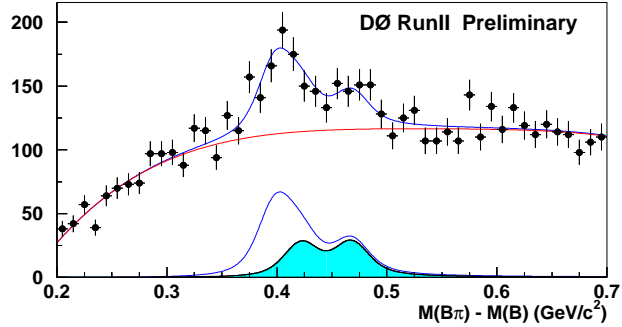}
\caption{Orbitally-excited P-wave B mesons, as reported by D0. 
Fitted contributions from 
B$_1\to$B$\pi$,
B$_2^* \to$B$\pi$ and 
B$_2^* \to$B$^*\pi$ are shown \cite{D0_Bnote}.}
\end{figure}

At the previous meeting HADRON03, the spectroscopy of the D$_s$ sector was a 
topic of great interest, due to the discovery of the surprisingly light and 
narrow states D$_{s0}^*(2317)$ and D$_{s1}(2460)$. Here we have heard 
relatively few new experimental results on these states. The most interesting 
new result may be their observation in B decays by Belle, reported here by Bianco 
\cite{Bianco}. The D$_{s0}^*(2317)$ is seen in the process 
B$^0 \to {\rm D}_{s0}^{*+}(2317) {\rm K}^-$, 
D$_{s0}^{*+}(2317)\to {\rm D}_s^+ \pi^0$, and the D$_{s1}^+(2460)$ 
may have been seen (at a much lower rate) in a similar chain to 
${\rm D}_s^+ \gamma$. The reported branching fractions are
\begin{equation}
B_{{\rm B}^0 \to {\rm D}_{s0}^{*+}(2317) {\rm K}^-} 
\cdot
B_{{\rm D}_{s0}^{*+}(2317) \to {\rm D}_s^+ \pi^0} 
= 4.4 \pm 0.8 \pm 0.6 \pm 1.1 \cdot 10^{-5},
\label{B_2317_BF}
\end{equation}
\begin{equation}
B_{{\rm B}^0 \to {\rm D}_{s1}^{*+}(2460) {\rm K}^-}
\cdot
B_{{\rm D}_{s1}^{*+}(2460) \to {\rm D}_s^+ \gamma}
= 0.53 \pm 0.20 \pm 0.6 {+0.16 \atop -0.15} \cdot 10^{-5}.
\label{B_2460_BF1}
\end{equation}
The latter branching fraction produce is also quoted as
$ <  0.86 \cdot 10^{-5}, \ \ 90\% \ c.l.$

Although it is very interesting to see the narrow D$_{sJ}$ mesons in
B decays, uncertainty regarding the production mechanism makes the interpretation
of these numbers in terms of the internal structure of the D$_{sJ}$ states 
somewhat problematic. 

\section{Theory at HADRON05}

There were just four plenary talks at HADRON05 by theorists specifically devoted
to theory. Since these contributions appear in the proceedings, here I will 
simply cite the topics discussed, and will consider one of these talks, which
addresses a long-standing problem in hadron physics, in more detail.
The theory plenary talks were given by Brambilla  
(effective field theories), Brodsky  
(anti deSitter space and conformal field theory), 
Fodor (recent developments in lattice QCD), and 
van~Beveren (``complex meson spectroscopy'').  

Brambilla \cite{Brambilla} 
discussed recent work on the derivation of effective Lagrangians 
for heavy-quark mesons (treated as fields) and gluons. This approach was 
cited as having many potential applications in the description of the physics
of heavy quark hadrons, including meson and baryon spectroscopy, radiative widths, 
determination of $m_c$, $m_b$ and $\alpha_s$, and (in the pure glue sector) the 
spectrum of glueballs.

Brodsky \cite{Brodsky} 
gave a remarkable talk on connections between string theory, Maldecena
duality, anti deSitter space, conformal field theory and hadrons. 
These connections
suggested a generalization of the MIT bag model, which leads to light cone 
wavefunctions that allowed the evaluation of many properties of hadrons in the
hard scattering regime. Predictions of this ``template for QCD'' included results
for the spectrum of mesons, baryons and $gg$ glueballs, and the absence of $ggg$ 
C-negative ``odderon'' glueballs (in a classical approximation to this approach).

Fodor \cite{Fodor} 
reviewed aspects of lattice QCD, and concentrated on
recent developments in the simulation of QCD at nonzero temperature.
Impressive new results have been reported for properties of the QCD phase 
diagram and equation of state, as a result of developments in algorithms 
for treating systems at nonzero chemical potential. 
These results have potential applications to an unusually broad set of
topics, including for example heavy ion collisions at RHIC and the internal 
structure of neutron stars.

Finally, van~Beveren \cite{vanBeveren}
discussed the effect of coupling the discrete, stable hadron
basis states of the naive quark model to the two-hadron continuum. This program,
often referred to as ``unquenching the quark model'', is an extremely important 
topic in hadron physics. It is clear that coupling naive quark model states
(or quenched lattice QCD states)
to the continuum of real and virtual decay channels will induce many important effects, 
such as mass shifts, nonzero decay widths, and modified electromagnetic 
couplings due to continuum components in the state vector. 

The narrow D$_{sJ}$ mesons may be examples of the
importance of the coupling of quark model states to the continuum. These states
lie {\it ca.}~100~MeV below the expectations of the naive quark model, 
which does not include loop effects. The fact that these states are near 
DK and D$^*$K thresholds and have very large S-wave couplings
to these channels suggests that their anomalously low masses may be due to 
mass shifts that result from virtual meson loops \cite{Hwang:2004cd}.    

The effects of virtual hadron loops on hadron properties have been
studied in detail only in a few specific cases, usually with 
a severely truncated set of continuum channels. 
Evaluation of these loop effects 
is complicated by the fact that the general three-hadron 
decay vertex is not well established, and several different models of this 
coupling have been proposed. Two specific forms that have seen wide 
application are the $^3$P$_0$ model of Micu \cite{Micu:1968mk} (which was 
used here by van~Beveren, and by most studies of loop effects) and 
a linear vector potential pair-production model, which was 
used by the Cornell group in their study of the effect of hadron loops 
on charmonium states \cite{Eichten:1979ms}. Application of the $^3$P$_0$ 
decay model to loop effects in $c\bar c$ suggests that the mass shifts are 
typically 100s of MeV \cite{Barnes:2004fs,Kalashnikova:2005ui}, much larger 
than the {\it ca.}~10~MeV quoted for the Cornell decay model 
\cite{Eichten:2004uh}. Distinguishing between these apparently very different 
predictions is complicated by the fact that the $^3$P$_0$ model predicts 
large {\it overall} mass shifts, but much smaller observable {\it differences} 
in mass shifts between states. 

These $^3$P$_0$ and Cornell decay models give quite different predictions for
some observables, such as the famous D/S ratio in $b_1\to \omega \pi$. 
Comparison of the different predictions with experimentally 
well established decay amplitudes should allow the development of a realistic 
working model of the crucial three-hadron vertices in future. 
The predicted mass shifts and other loop effects will clearly depend 
strongly on the model used to describe this coupling.

In their contribution to HADRON05, van~Beveren {\it et al.} describe a 
T-matrix formalism for iterating the effect of these meson loops, and 
apply it to the calculation of $2\to 2$ scattering amplitudes. 
Their results appear to be successful in describing many puzzling aspects of 
meson spectroscopy, such as the low-mass ``$\sigma$'' and ``$\kappa$''
scalar enhancements and the masses of the narrow D$_{sJ}$ states 
\cite{vanBeveren:2005ha}. 
This type of model shows considerable promise in attempts to understand 
the importance of hadron loops in the physics of hadrons, and in answering such
long-standing questions as why the valence quark model works as well as it does,
and under what circumstances it should fail.  

\section{O Pentaquark Morreu!}

In Elton Smith's presentation \cite{Smith} we learned that
Jefferson Lab has now accumulated considerable data on  
photoproduction of the kaon nucleon system (from both proton and deuteron targets),
so that they now have much better statistics than in the   
data originally cited as evidence for a narrow pentaquark 
\cite{Nakano:2003qx,Stepanyan:2003qr}. There is no evidence for any such 
state in this new high-statistics data \cite{Battaglieri:2005er}.
One may compare
the nK$^+$ mass distribution originally reported by 
the CLAS Collaboration in the reaction 
$\gamma d \to {\rm K}^+ {\rm K}^- {\rm pn}$ 
in Fig.4 of Ref.\cite{Stepanyan:2003qr}
(our figure~7) to the new, high-statistics proton data shown here by Smith
(our figure~8). The previously claimed $5 \sigma$ peak at 
$1.542\pm 0.005$~GeV has disappeared.
Thus, ``The pentaquark is dead!''

\begin{figure}[ht]
\includegraphics[height=.3\textheight]{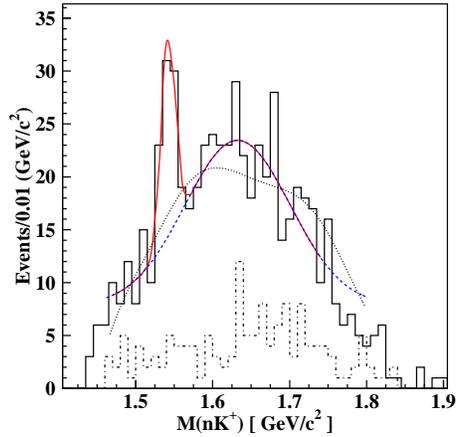}
\caption{The flavor-exotic pentaquark state previously reported by
the CLAS Collaboration in the nK$^+$ invariant mass distribution 
near 1540~MeV, in the reaction $\gamma d \to {\rm K}^+ {\rm K}^- {\rm pn}$
\cite{Stepanyan:2003qr}.}
\end{figure}

\begin{figure}[h]
\includegraphics[height=.3\textheight]{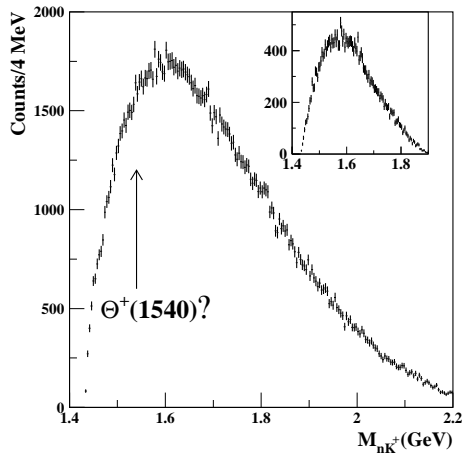}
\caption{The corresponding nK$^+$ invariant mass distribution from CLAS for
photoproduction on protons, presented at HADRON05 by 
Smith \cite{Smith}. Neither this nor the deuteron data show evidence of a
narrow pentaquark.}
\end{figure}

Since we have heard many experimental reports of observations of 
this or other related pentaquark states over the previous two years, 
and many theorists have claimed to understand this now vanished 
signal, it will certainly be of interest to revisit 
the pentaquark saga in future and ask just what mistakes were made, 
and what lessons we have learned as a result.
Here I will discuss one important issue in the theory of multiquark systems
generally; experimental problems with the pentaquark in particular
were reviewed recently by Dzierba {\it et al.} \cite{Dzierba:2004db}.

It is both interesting and unsettling that we have traveled 
this dead-end multiquark road many times before. 
The repeated claims of narrow multiquark resonances in 
low-statistics data sets accompanied by spurious theoretical support 
was such a problem in the 1970s that Isgur in 1985 wrote of 
``The Multiquark Fiasco'' in a series of Schladming lectures
\cite{Isgur:1985vp}. The discussion of multiquarks in 
this reference should be required reading for anyone working 
on the pentaquark, or on any multiquark resonance.

The conceptual problem with multiquarks was primarily one of 
understanding strong decays. Most models of multiquark systems simply 
excluded open decay channels from consideration, and assumed that the entire 
multiquark system was confined into a single hadron. Since there are 
many degrees of freedom present, this type of model predicts a 
very rich spectrum of discrete levels, many of which have exotic 
quantum numbers. One famous example is the I=2 $u^2\bar d^2$ $0^{++}$ 
flavor-exotic scalar, which was predicted by the MIT bag
model to have a mass of about 1.2~GeV. A resonant state with these quantum numbers 
should be clearly visible in the I=2 $\pi\pi$ S-wave phase shift, but experimentally
there is no indication of any such resonance; one sees only repulsive scattering. 
Although the bag model predicts a flavor-exotic multiquark scalar meson at 1.2~GeV, 
none is observed. What has gone wrong?

\begin{figure}[h]
\includegraphics[height=.15\textheight]{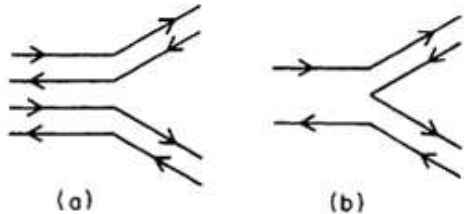}
\caption{Quark line diagrams showing (a) the spontaneous 
dissociation ``fall-apart decay'' of a multiquark meson into two color singlets, 
versus 
(b) the decay of a $q\bar q$ meson through $q\bar q$ pair production 
\cite{Jaffe:1975fd}.}
\end{figure}

The crucial mistake appears to have been the neglect of strong decay couplings, 
specifically the assumption by most models of a fully confined multiquark system. 
In a (perhaps oversimplified) quark model picture one can see that
the decays of multiquark states are fundamentally different from those of conventional 
$q\bar q$ or $qqq$ hadrons, which are the minimal color singlets. Decay
of a $q\bar q$ resonance into two $q\bar q$ mesons (for example) requires 
a decay interaction, specifically $q\bar q$ pair production.
In contrast, a multiquark system such as a $q^2\bar q^2$ meson or a 
$q^4\bar q$ pentaquark can simply spontaneously dissociate ``fall apart''
into two separate minimal color singlets, without a decay interaction. 
These two very different decay processes are shown in figure~9. 

Although the hadron community has absorbed this difference as the ``folklore'' 
that multiquark states are expected to be very broad,
the difference in decays can be much more drastic; unless there is some form of
fission barrier, a multiquark system above its strong fall-apart decay 
threshold will not exist as a resonance at all. This was appreciated in 
the late 1970s, and is the reason for Jaffe's statements \cite{Jaf77}
regarding his table of bag model dibaryon ($q^6$) multiquarks:
``Most of the "states" listed in the table probably do not correspond
to particles or resonances.'' The problem is that
``The object is classically unstable against small deformations leading 
to fission into separate n and p bags. [...] Quantizing about the six-quark object
would lead to instabilities analogous to those in a field theory with negative 
squared mass.'' In other words, these are not metastable states; 
the fall-apart effect implies that they do not exist as resonances. 

Since the proposed pentaquark was above the mass of a kaon and a nucleon, it too
has a fall-apart mode, and should not exist as a resonance barring 
exceptional circumstances such as a fission barrier. Alas this lesson appears
to have been forgotten, and multiquark models that did not consider fall-apart 
into the KN continuum were accepted as justification for the existence of
such states with undue credulity. 
\eject

\begin{figure}[h]
\includegraphics[height=.25\textheight]{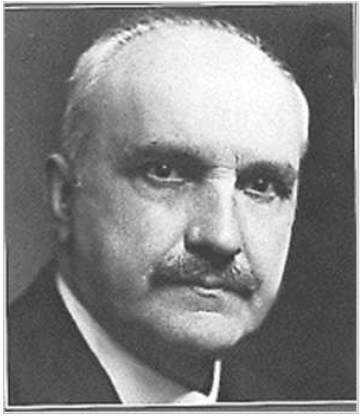}
\caption{
Jorge Agust\'in Nicol\'as Ruiz de Santayana y Borr\'as.
Professor of Philosophy, Harvard University, 1889-1912.}
\end{figure}

\vskip 0.5cm
Santayana's familiar quote unfortunately appears relevant to the pentaquark 
saga: ``Those who cannot remember the past are condemned to repeat it.'' 
\cite{San1}. A less well-known observation by Santayana, which is also relevant 
to the pentaquark story, is that ``Scepticism is the chastity of the intellect, 
and it is shameful to surrender it too soon or to the first comer.'' \cite{San2}

\subsection{Our moment of Zen}

``The story of the pentaquark shows how poorly we understand QCD.''
\cite{Wilczek}


\begin{theacknowledgments}
It is a pleasure to thank Alberto Reis and colleagues at CBPF 
for their kind invitation to present this summary,
and for the opportunity to discuss recent 
developments in hadron physics with my fellow participants. 
I also gratefully acknowledge useful communications from S.Bianco, 
S.Godfrey, J.Pursley, E.Smith and E.Swanson regarding aspects of the material 
discussed here. This research was supported in part by the U.S. National 
Science Foundation through grant NSF-PHY-0244786 at the University of Tennessee,
and the U.S. Department of Energy under contract
DE-AC05-00OR22725 at Oak Ridge National Laboratory.
\end{theacknowledgments}

\bibliographystyle{aipprocl} 

\eject

\end{document}